# Relative incapacitation contributions of pressure wave and wound channel in the Marshall and Sanow data set

Michael Courtney, PhD
Ballistics Testing Group, P.O. Box 24, West Point, NY 10996
Michael_Courtney@alum.mit.edu

Amy Courtney, PhD
Department of Physics, United States Military Academy, West Point, NY 10996
Amy_Courtney@post.harvard.edu

*Abstract:*
The Marshall and Sanow data set is the largest and most comprehensive data set available quantifying handgun bullet effectiveness in humans. This article presents an empirical model for relative incapacitation probability in humans hit in the thoracic cavity by handgun bullets. The model is constructed by employing the hypothesis that the wound channel and ballistic pressure wave effects each have an associated independent probability of incapacitation. Combining models for these two independent probabilities using the elementary rules of probability and performing a least-squares fit to the Marshall and Sanow data provides an empirical model with only two adjustable parameters for modeling bullet effectiveness with a standard error of 5.6% and a correlation coefficient *R = 0.939*. This supports the hypothesis that wound channel and pressure wave effects are independent (within the experimental error), and it also allows assignment of the relative contribution of each effect for a given handgun load. This model also gives the expected limiting behavior in the cases of very small and very large variables (wound channel and pressure wave), as well as for incapacitation by rifle and shotgun projectiles.
*Originally submitted 13 December 2006. Revised version submitted 1 August 2007.*

**I.     Introduction: Formulating the Pressure Wave Hypothesis**

Selecting service caliber handgun loads with the greatest potential for rapid incapacitation of violent criminal or terrorist attackers is of great interest in the law enforcement community. It is generally agreed upon that bullet design plays at least as important a role in bullet effectiveness as the cartridge from which it is fired. However, it is still widely debated whether the only contributing factors to the effectiveness of different designs are the volume of crushed tissue and penetration depth [FAC96a, PAT89].

Crushed tissue volume and penetration contribute to bullet effectiveness through the physiological consequences of blood loss. Over the years, other mechanisms such as hydrostatic shock, energy dump, hydraulic reaction, and the temporary stretch cavity have been suggested [PAT89]. Authors who suggest these mechanisms usually have something in mind more or less related to a ballistic pressure wave. However, other than temporary cavitation, these suggested mechanisms have largely eluded clear definitions and unambiguous evidence.

The view that the crushed tissue volume (the permanent cavity) [FAC96a] is the only reliable contributor to incapacitation (for handgun bullet placements which do not hit the central nervous system or supporting bone structure) relies on the unproven premise that easily detectable wounding[1] is necessary for an effect to contribute to incapacitation.

Even though it has been shown that most handgun bullets do not produce a temporary stretch cavity (the volume of tissue temporarily pushed out of the way by a passing bullet) large enough to stretch most vital tissue beyond the elastic limit, it remains an unproven premise that exceeding the elastic limit is necessary for correlating increased incapacitation with parameters determined from ballistic gelatin.

Because it has been shown that projectiles can cause remote neurological damage that requires microscopic examination or sensitive biochemical tests to detect [SHS90a, SHS90b, WWZ04], handgun bullet effectiveness should be studied without depending on the premise that the only contributors to incapacitation result in wounding that is easily observable to the trauma surgeon or medical examiner. One way to do this is to separate the issue of incapacitation from wounding by directly considering incapacitation and attempting to correlate observed metrics of incapacitation with potential causal agents. This was the approach of Marshall and Sanow in compiling their data [MAS92, MAS96, MAS01]. In contrast, work considering wounding or wound trauma as a valid incapacitation metric depends strongly on the unproven presupposition that only easily detectable wounding contributes to *rapid* incapacitation.

---

[1] By this we mean wounding that would typically be detected by a trauma surgeon or medical examiner.

For handgun loads that produce relatively large pressure waves, there is no published data showing that observed measures of rapid incapacitation are correlated only with the volume of crushed tissue or observed wounding. The conclusion that incapacitation is only caused by volume of crushed tissue or observed wounding is not founded upon repeatable experiments or carefully compiled data, but relies on expert opinion[2] [FAC87a, FAC96a, PAT89, MAC94] and exaggerated efforts to discredit data suggesting other causal agents [FAC91a, COC06a and references therein].

One step in the formation of the pressure wave hypothesis is the fact that most wound ballistics experts agree that a strongly energy-dependent factor such as the temporary stretch cavity plays a vital role in incapacitation via rifle bullets [FAC99a]. Frank Chamberlin observed pressure wave effects prior to World War II, and should probably be considered the father of the pressure wave hypothesis [CHA66]. It is not obvious that this energy dependence disappears as the energy-dependent parameter is lowered from rifle levels to handgun levels. The most powerful service caliber handgun loads only have 20-30% less impact energy than the .223 Remington.[3] In addition, since pressure wave magnitude depends on the local rate of energy transfer, some service caliber handgun loads have peak pressure wave magnitudes that are larger than some of the more deeply penetrating rifle loads.

A second step in the formation of the pressure wave hypothesis is our repeated observations from sport hunting that one only needs a handgun bullet which expands to 0.6" or 0.7" diameter to incapacitate a deer as quickly (with a shot to the center of the chest) as an archery broadhead of 1-3/8" to 1-1/2" cutting diameter. If the only important parameter was the amount of major blood vessels severed to create blood loss, then it stands to reason that a bullet would have to expand to a diameter comparable to that of the broadhead to have a comparable effect.

Yet with full broadside penetration near the center of the chest, hunting handgun bullets that reliably expand to only 0.6" or 0.7" in the typical deer hunting cartridges will reliably incapacitate deer at least as quickly (on average) as a hunting broadhead of 1-3/8" to 1-1/2" cutting diameter. If handgun bullets produce an effect in deer that provides more rapid incapacitation than bleeding effects alone, then it stands to reason that this effect may be present in similarly-sized mammals such as humans.

A third step in the formation of the pressure wave hypothesis is the idea that blood pressure drop due to internal bleeding takes about 5 seconds to have its effect in a best case scenario. Newgard describes this idea well [NEW92]:

*For an average 70 kg (155 lb.) male the cardiac output will be 5.5 liters (~1.4 gallons) per minute. His blood volume will be 60 ml per kg (0.92 fl. oz. per lb.) or 4200 ml (~1.1 gallons). Assuming his cardiac output can double under stress (as his heart beats faster and with greater force), his aortic blood flow can reach 11 liters (~2.8 gallons) per minute. If one assumes a wound that totally severs the thoracic aorta, then it would take 4.6 seconds to lose 20% of his blood volume from one point of injury. This is the minimum time in which a person could lose 20% of his blood volume.*

These theoretical ideas are confirmed by many observations of deer almost always taking 5-10 seconds to fall with any broadside archery shot hit through the center of the chest. In contrast, we have observed numerous deer drop in under 5 seconds when hit by handgun bullets creating pressure waves at the larger end of the spectrum [COC06d]. Likewise, events of apparently involuntary incapacitation in under 5 seconds are repeatedly reported in humans for handgun shots which fail to hit the CNS or supporting bone structure.

The Strasbourg tests employed a pressure sensor inserted into the carotid artery of live unanaesthetized goats [STR93]. These tests directly suggest that an internal pressure wave created by the interaction of the bullet and tissue can contribute to rapid incapacitation and can incapacitate more quickly than the crush cavity/blood loss mechanism alone:

*In a substantial number of cases, the subject was incapacitated almost instantly. Each time this occurred,*

---

[2] We agree with these authors on many points, especially their correlations of bullet parameters and gelatin measurements with observed wounding. Our scientific disagreement with their work is whether easily observed wounding alone is well correlated with *rapid* incapacitation. To the extent that they don't claim that it is, we are not saying that they are wrong as much as their work is incomplete. (Their claim is that wounding is well correlated with reliable *eventual* incapacitation. Our concern is *rapid* incapacitation.) We disagree only with an untested hypothesis for which they offer no data.

[3] The most energetic 10mm loads provide over 750 ft-lbs of energy. In an M4, the SS109 round provides 1026 ft-lbs at 50 yards and only 686 ft-lbs at 200 yards. Consequently, studying how incapacitation depends on pressure wave magnitude (related to energy) is also relevant for understanding decreasing .223 effectiveness as range is increased or barrel length is shortened.

*between two and five pressure spike tracings of high amplitude and short duration were found which immediately preceded and matched corresponding, diffused, or flattened lines (EEG tracings). Normally, the time lag between the first pressure spike and the beginning of slowed or flattened lines was between 30 and 40 milliseconds (although there were several cases where this delay lasted as long as 80 milliseconds)…The taller pressure spike tracings always preceded the slowed or flat line tracing…The initial spikes had to be of a certain height in order for the animal to collapse immediately.*

In contrast to the Strasbourg result, there have been some arguments against the pressure wave hypothesis. Claims to disprove the hypothesis of pressure wave incapacitation effects usually contain one or more of the following flaws [PAT89, FAC96a, MAC94]:

1. *Considering velocity ranges*
It is fallacious to consider pressure wave effects in terms of velocity ranges rather than pressure wave magnitude ranges. There is no velocity threshold where the pressure wave effects begin to turn on. There is a pressure wave magnitude threshold.

2. *Considering easily observable wounding*
It is fallacious to consider easily observable wounding rather than an observable measure of incapacitation. It is important to consider the possibility of incapacitation mechanisms that might not produce wounding that is easily detectable to a trauma surgeon or medical examiner. Consequently, to disprove the pressure wave hypothesis, one would have to observe and measure incapacitation directly rather than simply observing wounding after the fact. To our knowledge, there is no published data that fails to show a pressure wave contribution to incapacitation over the full range of peak pressure magnitudes produced by handgun loads.

3. *Considering only smaller pressure waves*
It is fallacious to show that the pressure wave does not make a significant contribution to incapacitation in a pressure wave regime considerably smaller than some available handgun loads. To compare different handgun loads, we consider the peak pressure on the edge of a 1" diameter circle centered on the axis of the wound channel. Handgun loads that produce 500 PSI at this point have pressure wave incapacitation effects that are difficult to discern (require a very large number of data points) with shots to the thoracic cavity. Handgun loads that produce over 1000 PSI at this point and penetrate at least 10" have pressure wave contributions that are relatively easy to discern (don't require as many data points).

4. *Confusing the concepts of unproven and disproven*
The fact that some previous studies make a weak case for the pressure wave mechanism means that the effect has not yet been proven in the published literature. This does not mean it has been disproven.

5. *Considering kinetic energy ranges*
It is fallacious to consider kinetic energy ranges rather than pressure wave ranges. It is not sufficient to show the absence of a pressure wave contribution to incapacitation with loads that produce a certain amount of kinetic energy. Disproving the pressure wave hypothesis for handgun bullets requires showing the absence of a pressure wave contribution to incapacitation with loads that produce 1500 PSI on the edge of a 1" diameter circle centered at the wound channel [COC06a].

6. *Considering only the "sonic pressure wave"*
The complete pressure wave (defined as the force per unit area that would be measured by a high-speed pressure transducer) must be considered.

Consequently, since there has been no convincing experimetal disproof of the pressure wave hypothesis, we should consider that the hypothesis of a pressure wave contribution to rapid incapacitation has substantial anecdotal support, and is well supported by the Strasbourg tests [STR93].

There are also a number of papers in the peer-reviewed journals suggesting ballistic pressure wave effects on wounding and incapacitation [GIK88, SHS87, SHS88, SHS89, SHS90a, SHS90b, WWZ04]. These papers are mainly concerned with velocities of rifle bullets, but the energy transfer and pressure waves produced are within the regime of pistol bullets.

Suneson et al. [SHS90a, SHS90b] report that peripheral high-energy missile hits cause pressure changes and damage to the nervous system. This experimental study on pigs used high-speed pressure transducers implanted in the thigh, abdomen, neck, carotid artery, and brain [SHS90a p 282]:

*A small transducer . . . mounted in the end of a polyethylene catheter . . . was implanted into the cerebral tissue in the left frontoparietal region about 10 mm from the midline and 5 mm beneath the brain surface through a drill hole (6-mm diameter).*

The sensor implanted in the brain measured pressure levels as high as 46 PSI and 50 PSI for pigs shot in the thigh as described. (See Figure 2C and 2D [SHS90a p 284].) The average peak positive pressure to the brain over the different test shots with that set-up was 34.7 PSI +/- 9.7 PSI. The error range does not represent uncertainty in individual measurements, but rather uncertainty in determination of the mean because of significant shot-to-shot variations in the pressure magnitude reaching the brain. (For a given local pressure magnitude, the distant pressure magnitude will show variation.)

Apneic periods were observed during the first few seconds after the shot, and both blood-brain and blood-nerve barrier damage were found. They concluded that "distant effects, likely to be caused by the oscillating high-frequency pressure waves, appear in the central nervous system after a high-energy missile extremity impact."

Martin Fackler published a reply to Suneson et al., asserting that "Shock Wave" is a myth [FAC91a]:

*In ascribing "local, regional, and distant injuries" to the sonic pressure wave, Suneson et al. have overlooked the effect of transmitted tissue movement from temporary cavitation. Since two distinct mechanisms are acting in the Suneson et. al experiment, one cannot arbitrarily assign any effects observed to only one of them.*

These and other criticisms of Suneson et al. are exaggerated [COC06a]. In fact, the results of Suneson et al. also find substantial agreement with later experiments in dogs conducted by an independent research group using a similar methodology [WWZ04]:

*The most prominent ultrastructural changes observed at 8 hours after impact were myelin deformation, axoplasmic shrinkage, microtubular diminution, and reactive changes of large neurons in the high-speed trauma group. These findings correspond well to the results of Suneson et al., and confirmed that distant effect exists in the central nervous system after a high-energy missile impact to an extremity. A high-frequency oscillating pressure wave with large amplitude and short duration was found in the brain after the extremity impact of a high-energy missile . . .*

There are a number of additional papers in the peer-reviewed journals studying the damage to the central nervous system caused by pressure wave effects [TLM05 and references therein]. Since their focus is on long-term effects, this research does not reach definitive conclusions regarding whether these pressure wave effects contribute to rapid incapacitation of humans. However, there is a growing body of evidence that pressure waves near 30 PSI can cause CNS damage that would usually be undetected by a trauma surgeon or medical examiner, but can be quantified with advanced neurological techniques.

There is also well-established evidence that pressure waves near 30 PSI applied to the brain can cause immediate incapacitation of laboratory animals. In a study applying a pressure wave directly to the brain via the lateral fluid percussion technique, Toth et al. [THG97] report both instantaneous incapacitation and cellular damage:

*The delivery of the pressure pulse was associated with brief (<120-200 sec), transient traumatic unconsciousness (as assessed by the duration of suppression of the righting reflex).*

One reasonably wonders what relevance these live animal experiments using the lateral fluid percussion technique to induce a pressure wave injury in laboratory animals have for understanding neurological pressure wave effects in humans. A 15-year review and evaluation of this question concluded [TLM05]:

*We conclude that the lateral fluid percussion brain injury model is an appropriate tool to study the cellular and mechanistic aspects of human traumatic brain injury…*

Consequently, there is significant support for the hypothesis of a pressure wave contribution to incapacitation and injury not only in anecdotal observations and incapacitation studies in goats, but also in well-established results of neurological experiments.

In addition, Chamberlin observed damage remote from the wound channel he ascribed to the hydraulic reaction of body fluids [CHA66]. Tikka et al. showed that ballistic pressure waves originating in the thigh reach the abdomen. Wounding and delayed recovery of peripheral nerves have been reported [LDL45, PGM46]. Pressure waves cause compound action potentials in peripheral nerves [WES82], and ballistic pressure waves have been shown capable of breaking bones [MYR88].

In summary, there are published results showing that a pressure wave can cause rapid neurological incapacitation and/or injury in goats [STR93], dogs [WWZ04], swine [SHS90a], several species of laboratory rats [THG97, TLM05], and even in whales [KNO03]. In many of these cases, detecting wounding requires advanced techniques such as electron microscopy, cellular analysis, EEG monitoring, and sophisticated chemical analysis. A study in humans also demonstrates the potential for pressure wave injury [OBW94].

Consequently, the assertion that incapacitation is only caused by wounding that is easily detectable to the trauma surgeon or medical examiner has been disproven. This opens the door to consider support for pressure wave contributions to incapacitation in studies observing incapacitation directly without concern for easily detectable wounding.

We state the pressure wave hypothesis as follows:

**Pressure wave hypothesis:**
*Other factors being equal, bullets producing larger pressure waves incapacitate more rapidly than bullets producing smaller pressure waves.*

The pressure wave hypothesis received direct experimental support in *Experimental Observations of Incapacitation via Ballistic Pressure Wave Without a Wound Channel,* [COC07a], and traumatic brain injury has been linked to the ballistic pressure wave in *Links between traumatic brain injury and ballistic pressure waves originating in the thoracic cavity and extremities* [COC07b].

## II. Marshall and Sanow methodology

The Marshall and Sanow study [MAS92, MAS96] collected data for a large number of shootings with a wide variety of handgun loads in an attempt to quantify the relative effectiveness of different handgun loads. The selection criteria included events where a criminal attacker was hit in the thoracic cavity with a single bullet and where the bullet and load could be accurately identified.

Events where the violent attacker ceased the attack without striking another blow or firing another shot were classified as "one shot stops." Events where the attacker delivered subsequent blows, fired subsequent shots, or ran more than 10 feet were classified as "failures." Events where the attacker retreated but covered less than 10 feet were classified as successful stops also. The "one shot stop" percentage was determined by dividing the total "stops" by the total number of events meeting the selection criteria (stops + failures).

The Marshall and Sanow methodology has been subject to some exaggerated criticisms [COC06a] that limit the accuracy of the data, but not the validity. Most epidemiological-type studies of this size and complexity represent compromises between the breadth of the selection criteria and the number of data points available. Other researchers might have made different choices about these trade-offs or differed in implementation details, but on the whole, the Marshall and Sanow "one-shot stop" (OSS) data set is a valuable contribution to understanding of incapacitation via handgun bullets.

## III. Correlation of Marshall and Sanow with Strasbourg Tests

The Strasbourg tests find significant experimental support in observations of remote nerve damage in pigs [SHS90a, SHS90b], experimental observations of suppressed EEG readings in swine [GIK88], experimental observations of remote nerve damage and suppressed EEG readings in dogs [WWZ04], experiments demonstrating incapacitation and neural injury due to a pressure wave in laboratory rats [TLM05], and experiments studying incapacitation in deer [COC06d].

With this degree of validation for the Strasbourg tests, correlation between the Strasbourg data and the Marshall and Sanow data set represents substantial experimental support for the validity and accuracy of the Marshall and Sanow OSS data.

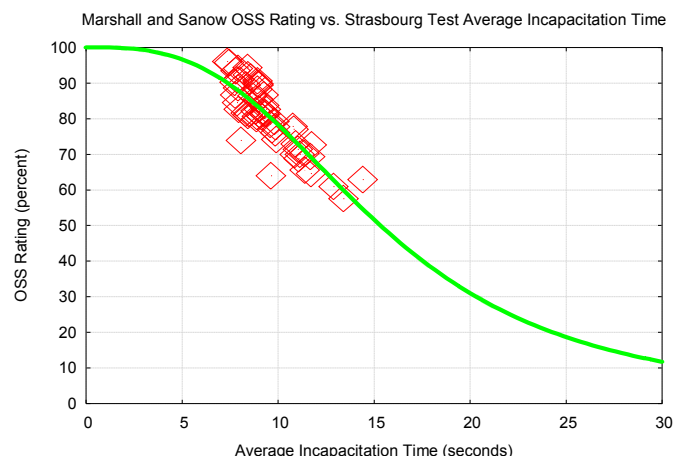

*Figure 1: Marshall and Sanow OSS rating plotted against Strasbourg test average incapacitation times for expanding handgun bullets along with best-fit model.*

The Marshall and Sanow OSS data [MAS96] are plotted against the Strasbourg average incapacitation times in Figure 1 for expanding bullets for which both the Strasbourg data and OSS data are available. Various model functions can be used to show an excellent correlation between the two data sets with a correlation coefficient R = 0.87. Because it has the expected limiting behavior for very large and very small average incapacitation times, we have chosen a model with the functional form:

$$OSS(t) = \frac{100}{1 + \left(\dfrac{t}{t_0}\right)^3}.$$

Performing a non-linear least squares fit gives a value for the adjustable parameter $t_0$ = 15.32 s. This model has a standard error of 4.87 which can be interpreted to mean that predicting Marshall and Sanow OSS ratings from the average incapacitation times of the goat tests will yield an expected accuracy of 4.87% in the predicted OSS rating. This level of agreement between the Marshall and Sanow data set and the Strasbourg results gives a high level of confidence in the validity of the Marshall and Sanow data set as an indicator of relative bullet effectiveness.

## IV. OSS Correlation with Bullet Performance Parameters

The Physics of the ballistic pressure wave is described elsewhere [COC06c]. The important result is that the peak pressure wave magnitude (in PSI) on the edge of a

1" diameter cylinder concentric with the bullet path can be estimated as

$$P_{1"} = \frac{5E}{d\pi},$$

where E is the kinetic energy (in ft-lbs) of the bullet at impact, and d is the penetration depth (in feet).

*Table 1: Correlation coefficients (R) and standard errors for OSS rating as a function of various bullet parameters. These correlations all used a linear least-squares fit to a third order polynomial.[4]*

| Bullet Parameter | R | Standard Error (%) |
|---|---|---|
| $P_{1"}$ | 0.845 | 8.45 |
| E | 0.871 | 7.99 |
| TSC | 0.861 | 8.27 |
| V | 0.530 | 13.79 |
| MV | 0.828 | 9.11 |
| $V_{PCC}$ | 0.760 | 10.56 |
| $A_{PCC}$ | 0.644 | 12.43 |

One can consider the Marshall and Sanow data and ask what bullet performance parameters might be used to most accurately predict the OSS rating. Performing linear least-squares fits (using a third order polynomial) to the OSS ratings as a function of various performance parameters yields the correlation coefficients and standard errors shown in Table 1.

The bullet performance parameters considered in the table are $P_{1"}$, the peak pressure wave magnitude defined above; E, the kinetic energy of the bullet; TSC, the volume of the temporary cavity; V, the bullet velocity; MV, the bullet mass times the velocity or the bullet momentum; $V_{pcc}$, the volume of the permanent crush cavity; and $A_{pcc}$, the surface area of the permanent crush cavity.

Unlike the Strasbourg Tests in which pressure wave magnitude is clearly the single bullet performance parameter most highly correlated with incapacitation [COC06c], the OSS rating is also highly correlated with energy, TSC, and momentum. This is because the OSS ratings include a wide variety of shot angles and placements, as well as a considerable voluntary component, whereas the shot placement and angle for the Strasbourg tests was fixed, and animal testing only models involuntary aspects of incapacitation in humans.

---
[4] Using various non-linear models can give the expected monotonic and limiting behaviors, but does not give a significantly smaller standard error or improved correlation. Third-order polynomials are adequate for answering which bullet performance parameters are more strongly related to the OSS rating.

Parameters more strongly related to energy (energy, TSC, and pressure wave magnitude) all show a higher degree of correlation than the parameters more strongly related to penetration and wound cavity.

A model including two independent variables demonstrates the greatest correlation with the OSS rating by including pressure wave magnitude and the surface area of the permanent crush cavity.

**V.     Empirical model of OSS incapacitation rating**

Steve Fuller was the first to confirm a high level of correlation between the Strasbourg results and the Marshall and Sanow OSS rating [MAS96, Ch 28]. He also developed empirical models to predict the OSS rating from parameters measured in ballistic gelatin. The accuracy of these predictive models was impressive, but objections were raised [MAC97] that the empirical models Fuller used did not give the proper limiting behavior in the cases of very large and small independent variables. One might also aspire to an empirical model that has fewer adjustable parameters (the Fuller best-fit model has five) and gives more physical insight into incapacitation mechanisms.

We have developed an alternative empirical model that addresses this objection and also offers insight into the basic scaling of incapacitation with pressure wave magnitude and permanent cavity size, as well as incorporating the laws of probability to support the hypothesis of independent pressure wave and crush cavity mechanisms.

Considering the hypothesis that the ballistic pressure wave is an incapacitation mechanism that works independently of the crush cavity, we can write the probability of the pressure wave (PW) mechanism *failing* to produce a one shot stop as:

$$P_{OSF}^{PW}(p) = \frac{1}{1 + \left(\frac{p}{p_0}\right)^{\frac{3}{2}}}$$

where $p$ is the peak pressure magnitude of a given load, and $p_0$ is the characteristic pressure magnitude that would produce a 50% failure rate in the limiting case of a very small permanent crush cavity. A possible experimental realization of this limiting case might be applying a pressure wave directly into the aorta via a catheter.

Likewise, we can write the independent one shot failure (OSF) probability of the blood pressure drop due to the permanent crush cavity (PCC) failing to produce a one shot stop as:

$$P_{OSF}^{PCC}(A) = \frac{1}{1+\left(\frac{A}{A_0}\right)^{\frac{3}{2}}}$$

where $A$ is the surface area[5] of the permanent crush cavity produced by a given load, and $A_0$ is the characteristic surface area that would produce a 50% failure rate in the limiting case of a very small pressure wave. An experimental realization of this limiting case might be creating a wound with a sharp instrument (such as an archery broadhead) that would initiate bleeding in a cylinder with a surface area of approximately $A$ while only creating a very small pressure wave.

The probability of a one shot *failure* when the independent mechanisms are both present is given by the product rule:

$$P_{OSF}^{Total}(p,A) = P_{OSF}^{PW}(p) P_{OSF}^{PCC}(A)$$

$$= \frac{1}{1+\left(\frac{p}{p_0}\right)^{\frac{3}{2}}} \times \frac{1}{1+\left(\frac{A}{A_0}\right)^{\frac{3}{2}}}.$$

The probability of a *successful* one shot stop is given by the complementary rule and is one minus the probability of failure:

$$P_{OSS}^{Total}(p,A) = 1 - P_{OSF}^{Total}(p,A) = 1 - P_{OSF}^{PW}(p) P_{OSF}^{PCC}(A)$$

$$= 1 - \left[\frac{1}{1+\left(\frac{p}{p_0}\right)^{\frac{3}{2}}} \times \frac{1}{1+\left(\frac{A}{A_0}\right)^{\frac{3}{2}}}\right].$$

The Marshall and Sanow OSS rating is then modeled by:

---

[5] Using a similar approach with wound volume produces a slightly less accurate result. The surface area of the permanent crush cavity is also preferred, because it stands to reason that the rate of blood loss should depend more on the surface area of wounded tissue than on the volume of wounded tissue.

$$OSS^{Total}(p,A) = 100\% \times P_{OSS}^{Total}(p,A)$$

$$= 100\% \times [1 - P_{OSF}^{PW}(p) P_{OSF}^{PCC}(A)].$$

An advantage of constructing a model for the OSS rating by considering the pressure wave and crush cavity as independent mechanisms is that the model can then be used to determine the relative contributions of these mechanisms for each load.

The one-shot failure (OSF) rating due to the pressure wave is:

$$OSF^{PW}(p) = 100\% \times P_{OSF}^{PW}(p).$$

By the complementary rule, the OSS rating due to the pressure wave is then

$$OSS^{PW}(p) = 100\% - OSF^{PW}(p)$$

$$= 100\% \times [1 - P_{OSF}^{PW}(p)].$$

In other words, this $OSS^{PW}$ rating describes the likely effectiveness of a projectile that applied an internal pressure wave with peak magnitude $p$, but only a very small crush cavity.

Likewise, the OSF rating due to the permanent crush cavity is:

$$OSF^{PCC}(A) = 100\% \times P_{OSF}^{PCC}(A).$$

By the complementary rule, the OSS rating due to the permanent crush cavity is then

$$OSS^{PCC}(A) = 100\% - OSF^{PCC}(A)$$

$$= 100\% \times [1 - P_{OSF}^{PCC}(A)].$$

In other words, this $OSS^{PCC}$ rating describes the likely effectiveness of a projectile that created a wound cavity with surface area $A$, but only a very small pressure wave.

A.         *Fitting the data*
The model described above allows a non-linear least squares fit to be performed on the Marshall and Sanow OSS data set using the peak pressure magnitude $p$ and the surface area *A* of the permanent crush cavity[6] as independent variables and the characteristic pressure $p_0$, and the characteristic area $A_0$ as adjustable parameters.

The characteristic pressure, $p_0$, is the peak pressure magnitude that would give an OSS rating of 50% for the case of very small crush cavity. The characteristic area, $A_0$, is the wound surface area that would give an OSS rating of 50% for the case of a very small pressure wave.

Since there are very few shot angles that are capable of creating incapacitating wounding at penetration depths beyond 20", penetration depths exceeding 20" were truncated to 20" when computing the surface area of the permanent cavity created by a given load. The best-fit parameters for 102 loads were $p_0$ = 377 PSI (+/- 30), $A_0$ = 28.5 Sq in (+/- 2.5), standard error = *6.2%*.

However, since loads producing less than 9.5" of penetration produce peak pressure waves at depths too shallow to have optimum effect, the 9 loads with penetration under 9.5" were removed from the data set.[7] The best fit parameters for the remaining 93 loads (listed in Appendix B):

**Best Fit Parameters:**
*$p_0$ = 339 PSI (+/- 27.86)*
*$A_0$ = 31.3 Sq in (+/- 3.102)*
**Standard Error = *5.6%***
*R = 0.939*

The characteristic pressure wave magnitude needed for a 50% OSS rating is smaller if one considers only loads penetrating at least *9.5"* deep. Shallow penetrating loads increase the peak pressure wave magnitude, but they do not place the wave deeply enough to have the optimal impact on vital organs. In subsequent discussions, we will use the parameters determined for the 93 handgun loads that produce at least *9.5"* of penetration in ballistic gelatin, and we will consider this to be our "Best Fit" model for the Marshall and Sanow OSS rating.

This result compares favorably with Steve Fuller's "Best Fit" model [MAS96, Ch 28] that has a standard error of 4.79% but has five adjustable parameters, three independent variables, gives little information about the independent mechanisms that might be involved, and does not demonstrate the expected limiting behavior at very large and very small values of the independent variables. In contrast, our model only uses two adjustable parameters (that have clear physical interpretations), two independent variables that are directly related to independent incapacitation mechanisms, and gives the expected limiting behavior at very small and very large pressure wave magnitude and crush cavity size.

B.  *Interpreting the model*
For a penetration depth of *12"*, we can interpret the characteristic area $A_0$ = *31.3* square inches in terms of an expanded bullet diameter of *0.83"*. Likewise, for *12"* of penetration, we can interpret the characteristic pressure $p_0$ = *339 PSI* in terms of a kinetic energy of *213 ft-lbs*.

The function,

$$OSS^{PW}(p) = 100\% \times P_{OSS}^{PW}(p) = 100\% \times \left[1 - P_{OSF}^{PW}(p)\right]$$

$$= 100\% \times \left[1 - \frac{1}{1 + \left(\frac{p}{p_0}\right)^{\frac{3}{2}}}\right]$$

is plotted for the best fit $p_0$ in Figure 2a. This function represents the independent contribution of the pressure wave to the OSS rating. One can consider this to be the OSS rating for a given pressure wave magnitude in the case of a very small permanent crush cavity.

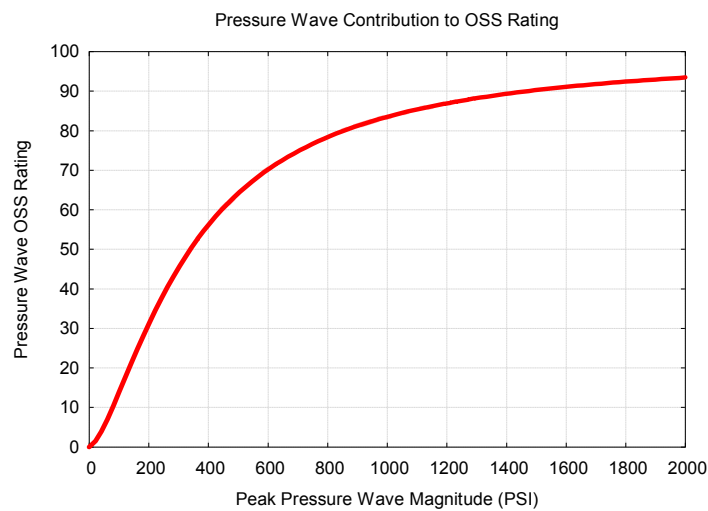

*Figure 2a: The best fit model pressure wave OSS rating is plotted as a function of peak pressure wave magnitude.*

---
[6] Here, we use the FBI model [PAT89] of the permanent crush cavity: a cylindrical region of tissue of diameter equal to the recovered final diameter of the bullet.

[7] Most loads with penetration under 9.5" are pre-fragmented specialty loads of little interest or use in law enforcement applications.

$OSS^{PW}(p)$ shows the expected limiting behavior, giving an OSS rating of 0% at 0 PSI and increasing in a monotonic manner to an OSS rating approaching 100% at very large pressure wave magnitudes. As expected, $OSS^{PW}(p)$ is equal to 50% at its characteristic pressure, $p_0 = 339$ PSI. $OSS^{PW}(p)$ crosses 70% at a peak pressure magnitude near 600 PSI (377 ft-lbs of energy for 12" penetration). $OSS^{PW}(p)$ is close to 80% at a peak pressure magnitude near 800 PSI (502 ft-lbs of energy for 12" penetration). However, the OSS rating of the pressure wave flattens considerably as the pressure rises above 800 PSI and (for very small permanent crush cavity) does not reach 90% until 1400 PSI (879 ft-lbs of energy for 12" penetration). In the absence of a significant crush cavity volume, the OSS rating for the pressure wave does not reach 95% until nearly 2500 PSI (1570 ft-lbs for 12" penetration).

The function,

$$OSS^{PCC}(A) = 100\% \times P_{OSS}^{PCC}(A) = 100\% \times \left[1 - P_{OSF}^{PCC}(A)\right]$$

$$= 100\% \times \left[1 - \frac{1}{1 + \left(\frac{A}{A_0}\right)^{\frac{3}{2}}}\right]$$

is plotted in Figure 2b. This represents the independent contribution of the permanent crush cavity to the OSS rating. This is the OSS rating for a given permanent crush cavity in the case of a very small pressure wave.

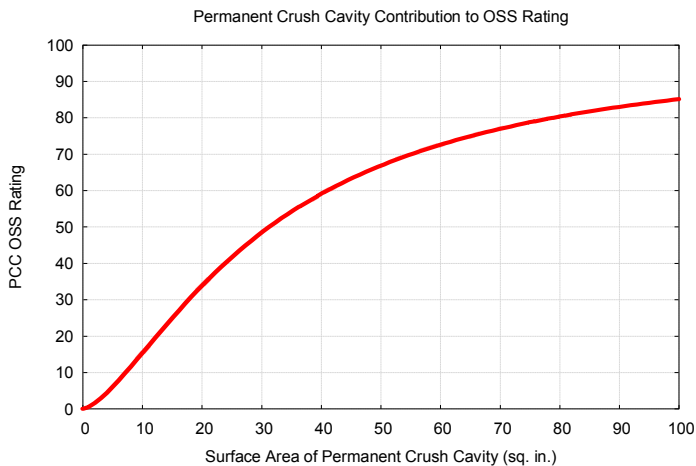

Figure 2b: The best-fit permanent crush cavity OSS rating is plotted as a function of surface area.

$OSS^{PCC}(A)$ shows the expected limiting behavior, giving an OSS rating of 0% at 0 sq. in. and slowly increasing in a monotonic manner to an OSS rating approaching 100% at very large permanent crush cavities. As expected, $OSS^{PCC}(A)$ is equal to 50% at its characteristic area, $A_0 = 31.3$ sq. in. $OSS^{PCC}(A)$ crosses 60% at an area near *40 sq. in.* This corresponds to a final diameter of *1.06"* for *12"* penetration or a final diameter of *0.80"* for *16"* of penetration. Very large expanded diameters seem necessary to create moderate levels of stopping power in the absence of pressure wave effects.

$OSS^{PCC}(A)$ is close to 70% at a surface area near *55 sq. in.* This corresponds to an expanded diameter of *1.45"* for *12"* of penetration and an expanded diameter of *1.09"* for *16"* of penetration. In the absence of pressure wave effects, the stopping power of relatively poorly performing 9mm loads requires expanded diameters comparable to archery broadheads or shotgun slugs.

The OSS rating of the permanent cavity flattens considerably as the PCC surface area rises above *50 sq. in.* and (for very small pressure waves) does not reach 90% until close to *136 sq. in.* This corresponds to an expanded diameter of *3.61"* for *12"* of penetration and an expanded diameter of *2.71"* for *16"* of penetration. According to this model, OSS ratings near 90% would not be possible with handgun bullets without pressure wave effects.

The combined OSS rating (including both pressure wave and permanent crush cavity effects) can be written as $OSS^{Total}(p, A) = 100\% \times P_{OSS}^{Total}(p, A) =$

$$100\% - 100\% \times \left[\frac{1}{1 + \left(\frac{p}{p_0}\right)^{\frac{3}{2}}} \times \frac{1}{1 + \left(\frac{A}{A_0}\right)^{\frac{3}{2}}}\right],$$

where $p_0$ and $A_0$ are the best fit characteristic parameters. This function is plotted in Figure 3 as a three dimensional graph along with the OSS data set.

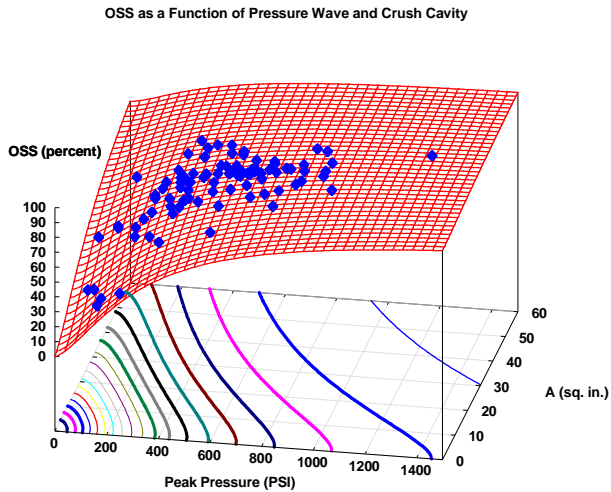

*Figure 3: The best-fit OSS model as a function of pressure wave and crush cavity surface area is plotted along with the OSS data.*

It can be challenging to understand the quantitative features of the model from the three dimensional graph, so the following discussion employs separate two dimensional graphs and lines of constant crush cavity area and pressure wave, respectively.

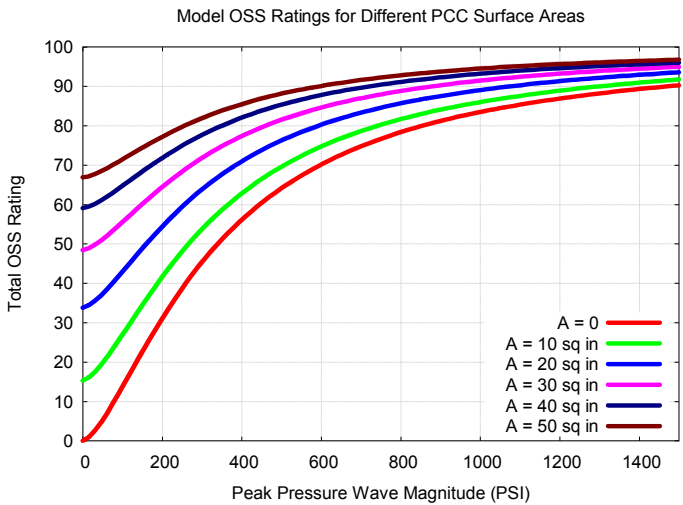

*Figure 4: Best-fit model OSS ratings plotted as a function of pressure wave magnitudes for a selection of crush cavity surface areas.*

Figure 4 shows a two dimensional graph of $OSS^{Total}(p,A)$, by plotting $OSS^{Total}(p, A=0)$, $OSS^{Total}(p, A=10)$, $OSS^{Total}(p, A=20)$, $OSS^{Total}(p, A=30)$, $OSS^{Total}(p, A=40)$, and $OSS^{Total}(p, A=50)$. As the crush cavity surface area increases, $OSS^{Total}(p,A)$ has a higher value for $p = 0$ PSI, ranging from 0% at *0 sq. in.* to close to 67% at *50 sq. in.* We can also make the following observations:.

- At *p = 200 PSI*, $OSS^{Total}(p,A)$ ranges from 30% at *A = 0 sq. in.* to 78% at *A = 50 sq. in.*
- At *p = 400 PSI*, $OSS^{Total}(p,A)$ ranges from 55% at *A = 0 sq. in.* to 85% at *A = 50 sq. in.*
- At *p = 800 PSI*, $OSS^{Total}(p,A)$ ranges from 78% at *A = 0 sq. in.* to 93% at *A = 50 sq. in.*
- At *p = 1200 PSI*, $OSS^{Total}(p,A)$ ranges from 87% at *A = 0 sq. in.* to 96% at *A = 50 sq. in.*

It is clear that as the peak pressure wave magnitude increases above *800 PSI*, the effect of relatively small differences in crush cavity volume become progressively less important.

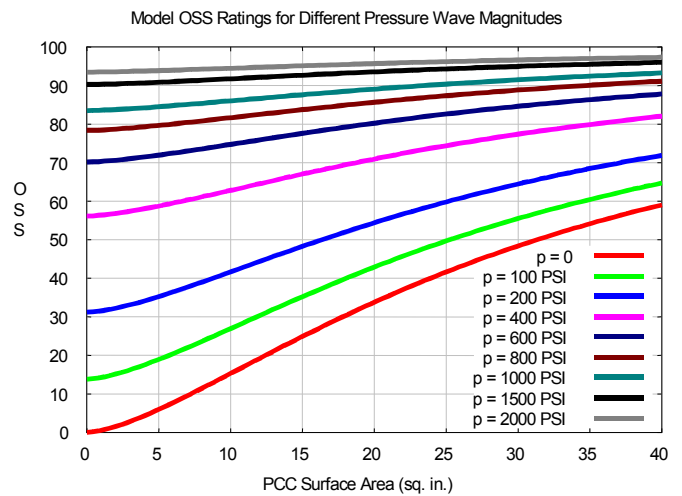

*Figure 5: Best-fit model OSS ratings plotted as a function of PCC surface area for a selection of peak pressure wave magnitudes.*

Figure 5 shows a two dimensional graph of $OSS^{Total}(p,A)$, by plotting $OSS^{Total}(p=0, A)$, $OSS^{Total}(p=100, A)$, $OSS^{Total}(p=200, A)$, $OSS^{Total}(p=400, A)$, etc., up to $OSS^{Total}(p=2000, A)$. As the pressure increases, $OSS^{Total}(p,A)$ has a higher value for *A = 0 sq. in.*, ranging from 0% at *0 PSI.* to close to 93% at *2000 PSI.* We also make the following observations:

- At *A = 10 sq. in.*, $OSS^{Total}(p,A)$ ranges from 15% at *p = 0 PSI* to 95% at *p = 2000 PSI.*
- At *A = 20 sq. in.*, $OSS^{Total}(p,A)$ ranges from 33% at *p = 0 PSI* to 96% at *p = 2000 PSI.*
- At *A = 30 sq. in.*, $OSS^{Total}(p,A)$ ranges from 48% at *p = 0 PSI* to 97% at *p = 2000 PSI.*
- At *A = 40 sq. in.*, $OSS^{Total}(p,A)$ ranges from 58% at *p = 0 PSI* to 97% at *p = 2000 PSI.*

At all PCC sizes produced by handgun bullets, the OSS rating increases significantly with increasing pressure.

Once again, we see that as the peak pressure wave magnitude increases above *800 PSI*, the effects of relatively small differences in crush cavity volume become progressively less important.

## VI. Interpretation and Conclusions

Appendix B contains a table giving the $OSS^{PW}$, $OSS^{PCC}$, and $OSS^{TOTAL}$ ratings for 93 different handgun loads. It is interesting to note how the independent pressure wave and crush cavity mechanisms can combine to produce total OSS ratings in different loads. For example, in the 9mm Win 147 grain JHP, $OSS^{PW}$ = 54.7% and $OSS^{PCC}$= 47.1%. These two independent mechanisms combine by the rules of probability to produce $OSS^{TOTAL}$ = 76.0%. (The M&S OSS rating is 74.1% for this load.)

Most loads have a pressure wave contribution to the OSS rating that is significantly higher than the crush cavity contribution to the OSS rating. For example, the 9mm Fed 124 HS +P+ has $OSS^{PW}$ = 69.3% and $OSS^{PCC}$= 46.1%. These two mechanisms combine to produce $OSS^{TOTAL}$ = 83.5%. (The M&S OSS rating is 85.5% for this load.) The .380 ACP Fed 90 HS has $OSS^{PW}$ = 52.6% and $OSS^{PCC}$= 32.3%. These two mechanisms combine to produce $OSS^{TOTAL}$ = 67.9%. (The M&S OSS rating is 69.0% for this load.)

The few loads for which the crush cavity contribution to the OSS rating exceeds the pressure wave contribution are loads that do not expand. For example, the Fed 158 grain semi-wad cutter (SWC) in .38 Special has $OSS^{PW}$ = 20.3% and $OSS^{PCC}$= 38.1%. These two mechanisms combine to produce $OSS^{TOTAL}$ = 50.6%. (The M&S OSS rating is 51.8% for this load.) Similarly, the Rem 230 grain FMJ in .45 ACP has $OSS^{PW}$ = 39.0% and $OSS^{PCC}$= 46.2%. These two mechanisms combine to produce $OSS^{TOTAL}$ = 67.2%. (The M&S OSS rating is 64.8% for this load.)

The standard error of the model is 5.6% which gives the typical level of accuracy when comparing the model to the actual M&S OSS ratings. Some loads have a noticeably larger deviation from the model than the standard error. We believe that there are several factors that can cause a load to exhibit a deviation from the model that is noticeably larger than the standard error.[8]

We believe that loads significantly over-perform (compared to the model) because:
- There are an insufficient number of data points to yield an accurate OSS estimate. The uncertainty in the M&S OSS rating is at least $0.5/N^{1/2}$ for N shooting events.

- Bullet fragmentation creates a pressure wave significantly larger than our estimate (which ignores fragmentation). Accounting for fragmentation is described in Appendix A.[9]
- Some circumstances lead to a given load having a higher percentage of upper thoracic cavity hits. These circumstances might include better training and discipline for departments using the load.

The Remington 125 grain JHP load in .357 Magnum is an example of an over-performing load. This load has model $OSS^{TOTAL}$ = 87.1% but an actual M&S OSS rating of 96.1%. This OSS rating includes 204 shooting events, so the uncertainty can be estimated as 3.5%. Accounting for fragmentation[10] raises the peak pressure magnitude estimate from 796.8 PSI to 1274.9 PSI, which raises the $OSS^{PW}$ from 78.2% to 87.9%. This increased $OSS^{PW}$ combines with an $OSS^{PCC}$ = 40.5% to create an increased $OSS^{TOTAL}$ = 92.8%. This does not completely account for the difference between the model and the M&S OSS rating of 96.1%, but it does bring them significantly closer.

Another over-achieving load is the Cor-Bon 135 grain JHP load in .40 S&W. This load has model $OSS^{TOTAL}$ = 88.1% but an actual M&S OSS rating of 95.8%. Accounting for fragmentation[11] raises the peak pressure magnitude estimate from 988.2 PSI to 1383.5 PSI, which raises the $OSS^{PW}$ from 83.2% to 89.1%. This increased $OSS^{PW}$ combines with an $OSS^{PCC}$ = 29.0% to create an increased $OSS^{TOTAL}$ = 92.3%. This does not completely account for the difference between the model and the M&S OSS rating of 95.8%, but it brings them closer.

Many loads that significantly under-perform our best-fit model are loads with large levels of recoil suggesting the possibility of a higher percentage of abdominal hits than occurring in the data set as a whole.[12] For example, the Win 210 grain JHP in .41 Magnum has an $OSS^{TOTAL}$ = 91.9% but only has M&S OSS rating of 82.4%. The

---

[8] Most loads that significantly over-perform or under-perform the predictions of our model also deviate significantly from all of Steve Fuller's models.

[9] We did not account for fragmentation in our initial estimates of pressure magnitude because we were unable to obtain data on the retained mass of many bullets. Many designs are no longer available for testing.

[10] As described in Appendix A, using a fragmentation percentage of 30% and an average fragment penetration depth of 33% of the overall penetration depth.

[11] As described in Appendix A, using a fragmentation percentage of 40% and an average fragment penetration depth of 50% of the overall penetration depth. The 135 grain Nosler JHP bullet sheds larger fragments than the 125 grain Remington JHP in .357 Magnum, so the fragments tend to penetrate more deeply.

[12] Marshall and Sanow make no mention of the ratio of abdominal to chest hits in their data set. Loads that have a higher percentage of abdominal hits would be expected to have a lower OSS rating in their data set.

Rem 210 grain JSP in .41 Magnum has an $OSS^{TOTAL}$ = 91.9% but the M&S OSS rating falls far short of this at 80.7%. The Win 225 grain Silvertip load in .45 Long Colt has an $OSS^{TOTAL}$ = 84.4% but the M&S OSS rating falls far short of this at 73.6%.

Beyond the ability to predict the OSS rating from bullet performance parameters, our model provides important insights into the relative contributions of the permanent crush cavity and ballistic pressure wave. To the degree of accuracy possible with this data set, the permanent crush and pressure wave mechanisms appear to be independent. For most JHP handgun loads, the pressure wave contribution to the OSS rating is significantly larger than the crush cavity contribution. Our model provides a way to quantify the trade-offs inherent when considering increasing one bullet performance parameter at the expense of another.

*A.       Cautions and Limits of Interpretation*
The M&S OSS rating is a relative measure of bullet effectiveness that does not predict the outcome of specific shooting events but rather compares the relative effectiveness of different handgun loads. Our model should be considered the same way. A OSS rating of 90% does NOT mean that a load has a 90% chance of stopping a deadly attack with a hit to the torso, it simply means that a load is likely to perform better than a load with an 80% OSS rating and worse than a load with a 96% OSS rating. Of course, the data set includes both voluntary and involuntary stops, so the model gives no indication of the real probability of stopping a particularly determined attacker.

Since the model was developed using only loads that penetrate at least 9.5", it should only be considered valid for loads that penetrate at least 9.5". Bullets that penetrate less will underperform predictions based on this model.

We've shown that an empirical model of the OSS rating of a load can be developed from parameters measured in ballistic gelatin. However, our model based on the Strasbourg tests makes more accurate OSS predictions for JHP handgun loads than our gelatin-based model. This suggests that live animal testing can be productively incorporated into testing of bullet effectiveness. Since our deer testing methodology [COC06d] is well correlated with the Strasbourg tests and deer are widely available, deer testing of service caliber handgun loads can be considered a valuable addition to gelatin testing.

The M&S OSS rating contains a significant component of voluntary incapacitation. The approach here folds all voluntary incapacitation into the pressure wave and permanent cavitation contributions. It would be possible to model the voluntary contribution to incapacitation as a third, independent incapacitation mechanism, combine the three independent probability functions by the rules of probability, and do a least-squares fit to the resulting composite model. This could estimate independent OSS contributions for voluntary, permanent crush, and pressure wave effects.

This approach can be generalized to any number of independent effects and points out the benefit of new technologies which would utilize new incapacitation mechanisms not currently employed with current technology.

In conclusion, this analysis shows that, all other factors being equal, bullets that produce pressure waves of greater magnitude incapacitate more rapidly than bullets that produce smaller pressure waves. The M&S OSS data convincingly supports the pressure wave hypothesis and allows (perhaps for the first time) human incapacitation to be modeled as a function of peak ballistic pressure wave magnitude.

*B.       Implications for Bullet Testing and Selection*
The success of our empirical model in describing bullet effectiveness suggests that it might be used for predicting bullet effectiveness before data from actual shooting events is available. Estimating an OSS rating from actual shootings requires at least 80 shooting events meeting the selection criteria to provide the expected level of accuracy offered by our model. Both the pressure wave magnitude and surface area of the permanent crush cavity can easily be determined by using a chronograph to measure bullet energy and results from shooting into calibrated 10% ballistic gelatin to determine final bullet diameter, penetration depth, and retained mass.

However, one should not be overly impressed by the propensity for shallow penetrating loads to produce larger pressure waves. Bullet selection criteria should first determine the required penetration depth for the given risk assessment and application, and only use pressure wave magnitude as a selection criterion for bullets that meet a minimum penetration requirement.

Reliable expansion, penetration, feeding, and functioning are all important aspects of load testing and selection. We do not advocate abandoning long-held aspects of the load testing and selection process, but it seems prudent to consider the pressure wave magnitude and the predicted OSS rating along with other factors.

*C.       Implications for Bullet Design*
The trend in bullet design over the last decade has drifted toward bullets with little fragmentation and a higher percentage of retained mass. Bullets that both fragment and meet minimum penetration requirements create larger pressure wave magnitudes and offer improved incapacitation potential.

In addition to moving toward designs which both penetrate and fragment reliably, we believe that the incapacitation potential of a bullet can be further improved by delaying expansion and fragmentation to a penetration depth of at least 4". This would place the peak pressure magnitude closer to vital organs.

Optimal use of a bullet's kinetic energy to produce pressure wave incapacitation suggests a bullet design that penetrates the first 4" or so prior to significant expansion or energy loss, and then rapidly expands and transfers a large percentage of its energy and 40% of its mass at penetration depths between 4-8" before continuing to penetrate to the depth desired for the application.

14

**Appendix A: Accounting for Fragmentation in Estimating the Peak Pressure Wave Magnitude**

If kinetic energy and penetration depth are equal, bullets that fragment create a larger pressure wave than bullets that retain 100% of their mass. This is because the average penetration depth is shorter than the maximum penetration depth. Recall that the average force with no mass loss is given by [COC06c]

$F_{ave} = E/d$,

where E is the kinetic energy and d is the maximum penetration depth.

If we consider the case of a bullet with some fraction, f, of mass lost to fragmentation, the fraction of retained mass is (1-f) and the average force is then given by

$F_{ave} = (1-f)E/d + f\,E/d_f$,

where $d_f$ is depth of the center of mass of the bullet fragments. In other words, $d_f$ is the average penetration depth of the fragments. Most fragments do not penetrate as deeply as the maximum penetration depth d, so that the average fragment penetration depth $d_f$ can be expressed as a fraction of the maximum penetration depth

$d_f = d/k$,

where k is greater than 1. Consequently, the average force becomes,

$F_{ave} = (1-f)E/d + f\,k\,E/d$.

This can be rewritten as

$F_{ave} = [1 + f(k-1)]E/d$.

So we see that the enhancement factor for the average force is $[1 + f(k-1)]$, where f is the fraction of lost mass, and k describes the relative penetration depth of the mass lost by fragmentation. If the mass lost by fragmentation penetrates ½ of the maximum penetration depth on average, k = 2, and the enhancement factor for the average force is (1+f). In other words, a 40% loss of mass increases the average force (and thus the pressure wave) by 40%.

If the mass lost by fragmentation penetrates ⅓ of the maximum penetration depth on average, k = 3, and the enhancement factor for the average force is (1+2f). In other words, a 40% loss of mass increases the average force (and thus the pressure wave) by 80%.

Consequently, bullets that fragment can create larger pressure waves than bullets that do not fragment but have the same kinetic energy and penetration depth. Most fragmenting bullets have an average fragment penetration depth of ⅓ to ½ of their maximum penetration depth, so that the pressure wave enhancement factor is between (1+f) and (1+2f).

**About the Authors**

*Amy Courtney* currently serves on the faculty of the United States Military Academy at West Point. She earned a MS in Biomedical Engineering from Harvard University and a PhD in Medical Engineering and Medical Physics from a joint Harvard/MIT program. She has taught Anatomy and Physiology as well as Physics. She has served as a research scientist at the Cleveland Clinic and Western Carolina University, as well as on the Biomedical Engineering faculty of The Ohio State University.

*Michael Courtney* earned a PhD in experimental Physics from the Massachusetts Institute of Technology. He has served as the Director of the Forensic Science Program at Western Carolina University and also been a Physics Professor, teaching Physics, Statistics, and Forensic Science. Michael and his wife, Amy, founded the Ballistics Testing Group in 2001 to study incapacitation ballistics and the reconstruction of shooting events. www.ballisticstestinggroup.org

Revision information:
*13 December 2006 to 1 August 2007:* Fixed typographical errors. Added Figure 3 (three dimensional figure of OSS model). Updated references, contact information, and biographies. Added references [CHA66], [LDL45], [PGM46], [MYR88], [WES82], [TCR82], [COC07a] and [COC07b].

## Appendix B: Data Table (M&S OSS data [MAS96])

| Cartridge | Load | $P_{max}$ (PSI) | OSS (M&S) | $P_{OSF}$ | $P_{OSS}^{PW}$ | $P_{OSS}^{PCC}$ | $P_{OSS}^{TOT}$ |
|---|---|---|---|---|---|---|---|
| .357Mag | Rem125JHP | 796.83 | 96.08 | 0.039 | 0.782 | 0.405 | 0.871 |
| .357Mag | Fed125JHP | 929.64 | 95.79 | 0.042 | 0.819 | 0.409 | 0.893 |
| .357Mag | CCI125JHP | 719.72 | 93.46 | 0.065 | 0.755 | 0.449 | 0.865 |
| .357Mag | Fed110JHP | 783.19 | 90.2 | 0.098 | 0.778 | 0.203 | 0.823 |
| .357Mag | Rem110JHP | 725.18 | 88.68 | 0.113 | 0.757 | 0.304 | 0.831 |
| .357Mag | Win125JHP | 832.51 | 86.67 | 0.133 | 0.793 | 0.420 | 0.880 |
| .357Mag | Win145ST | 716.00 | 84.52 | 0.155 | 0.754 | 0.474 | 0.871 |
| .357Mag | Rem125SJHP | 508.98 | 82.61 | 0.174 | 0.647 | 0.558 | 0.844 |
| .357Mag | Rem158SJHP | 537.88 | 81.58 | 0.184 | 0.666 | 0.482 | 0.827 |
| .357Mag | Fed158NY | 619.37 | 80.95 | 0.191 | 0.711 | 0.510 | 0.859 |
| .357Mag | CCI140JHP | 729.58 | 73.91 | 0.261 | 0.759 | 0.462 | 0.870 |
| .357Mag | Win158SWC | 392.70 | 72.45 | 0.276 | 0.554 | 0.381 | 0.724 |
| .357Mag | Rem158SWC | 371.62 | 67.61 | 0.324 | 0.534 | 0.381 | 0.711 |
| .380ACP | Fed90HS | 363.85 | 68.97 | 0.310 | 0.526 | 0.324 | 0.679 |
| .380ACP | FED90JHP | 265.31 | 68.81 | 0.312 | 0.409 | 0.273 | 0.570 |
| .380ACP | CCI88JHP | 220.24 | 57.59 | 0.424 | 0.343 | 0.325 | 0.557 |
| .380ACP | Rem88JHP(old) | 286.53 | 56.86 | 0.431 | 0.437 | 0.277 | 0.593 |
| .380ACP | Hor90XTP | 323.77 | 53.85 | 0.462 | 0.482 | 0.274 | 0.624 |
| .380ACP | Fed95FMJ | 216.87 | 51.38 | 0.486 | 0.338 | 0.325 | 0.553 |
| .38Sp | CB115+P+ | 494.92 | 82.68 | 0.173 | 0.638 | 0.459 | 0.804 |
| .38Sp | Win158LHP+P | 358.81 | 77.81 | 0.222 | 0.521 | 0.469 | 0.746 |
| .38Sp | Fed158LHP+P | 349.37 | 77.03 | 0.230 | 0.511 | 0.467 | 0.739 |
| .38Sp | Fed125JHP+P | 388.31 | 72.9 | 0.271 | 0.550 | 0.437 | 0.747 |
| .38Sp | Rem125SJHP+P | 404.90 | 72.64 | 0.274 | 0.566 | 0.395 | 0.737 |
| .38Sp | CCI125JHP+P | 338.38 | 70.27 | 0.297 | 0.499 | 0.494 | 0.746 |
| .38Sp | Rem158LHP+P | 358.81 | 69.93 | 0.301 | 0.521 | 0.504 | 0.762 |
| .38Sp | Rem95SJHP+P | 529.40 | 65.55 | 0.345 | 0.661 | 0.246 | 0.744 |
| .38Sp | Win125JHP+P | 367.24 | 64.62 | 0.354 | 0.529 | 0.447 | 0.740 |
| .38Sp | Fed158SWC | 136.44 | 51.8 | 0.482 | 0.203 | 0.381 | 0.507 |
| .38Sp | Fed158RNL | 136.44 | 50.66 | 0.493 | 0.203 | 0.381 | 0.507 |
| 9mm | CB115JHP | 626.87 | 90.63 | 0.094 | 0.715 | 0.410 | 0.832 |
| 9mm | Fed115JHP+P+ | 392.96 | 89.91 | 0.101 | 0.555 | 0.555 | 0.802 |
| 9mm | Rem115JHP+P+ | 687.37 | 89.47 | 0.105 | 0.742 | 0.393 | 0.843 |
| 9mm | Fed124HS+P+ | 584.47 | 85.51 | 0.145 | 0.693 | 0.461 | 0.835 |
| 9mm | Fed124NYLHP | 574.73 | 83.68 | 0.163 | 0.688 | 0.366 | 0.802 |
| 9mm | Fed115JHP | 657.12 | 81.73 | 0.183 | 0.729 | 0.345 | 0.823 |
| 9mm | Rem115JHP | 449.23 | 81.44 | 0.186 | 0.604 | 0.462 | 0.787 |
| 9mm | Fed124HS | 493.24 | 81.33 | 0.187 | 0.637 | 0.420 | 0.789 |
| 9mm | CCI115JHP | 501.07 | 79.26 | 0.207 | 0.642 | 0.385 | 0.780 |
| 9mm | Fed147HS | 436.81 | 78.42 | 0.216 | 0.593 | 0.468 | 0.784 |
| 9mm | Fed147JHP9MS2 | 328.98 | 77.78 | 0.222 | 0.488 | 0.498 | 0.743 |
| 9mm | Win147RSXT | 386.88 | 76.92 | 0.231 | 0.549 | 0.494 | 0.772 |
| 9mm | Fed147JHP9MS1 | 372.43 | 76 | 0.240 | 0.535 | 0.465 | 0.751 |
| 9mm | Win147JHP | 384.45 | 74.14 | 0.259 | 0.547 | 0.471 | 0.760 |
| 9mm | Win115FMJ | 265.87 | 62.89 | 0.371 | 0.409 | 0.381 | 0.634 |

| Cartridge | Load | $P_{max}$ (PSI) | OSS (M&S) | $P_{OSF}$ | $P_{OSS}^{PW}$ | $P_{OSS}^{PCC}$ | $P_{OSS}^{TOT}$ |
|---|---|---|---|---|---|---|---|
| .40S&W | CB135JHP(N) | 988.25 | 95.83 | 0.042 | 0.832 | 0.290 | 0.881 |
| .40S&W | Fed155JHP | 713.15 | 94.12 | 0.059 | 0.753 | 0.409 | 0.854 |
| .40S&W | Rem165GS | 772.05 | 93.75 | 0.063 | 0.774 | 0.426 | 0.870 |
| .40S&W | Fed155HS | 643.44 | 92.86 | 0.071 | 0.723 | 0.464 | 0.852 |
| .40S&W | C-B150JHP | 654.93 | 92.31 | 0.077 | 0.728 | 0.405 | 0.838 |
| .40S&W | Win155ST | 707.49 | 90.91 | 0.091 | 0.751 | 0.480 | 0.870 |
| .40S&W | Fed180HS | 459.73 | 89.47 | 0.105 | 0.612 | 0.546 | 0.824 |
| .40S&W | CB180JHP+P | 561.60 | 86.36 | 0.136 | 0.680 | 0.546 | 0.855 |
| .40S&W | BH180XTP | 420.48 | 84.78 | 0.152 | 0.580 | 0.520 | 0.798 |
| .40S&W | Eld180SF | 617.64 | 83.33 | 0.167 | 0.711 | 0.462 | 0.844 |
| .40S&W | Win180JHP | 533.81 | 80.95 | 0.191 | 0.664 | 0.476 | 0.824 |
| .40S&W | Win180RSXT | 576.00 | 80 | 0.200 | 0.689 | 0.477 | 0.837 |
| .40S&W | Win180FMJ | 165.50 | 70.59 | 0.294 | 0.254 | 0.502 | 0.628 |
| 10mmMV | Win180JHP | 548.42 | 81.82 | 0.182 | 0.673 | 0.510 | 0.840 |
| 10mmMV | Fed180JHP | 489.02 | 81.48 | 0.185 | 0.634 | 0.509 | 0.820 |
| 10mmMV | Rem180JHP | 566.70 | 80.65 | 0.194 | 0.683 | 0.474 | 0.834 |
| .45ACP | Fed230HS | 587.39 | 94.37 | 0.056 | 0.695 | 0.477 | 0.840 |
| .45ACP | Rem230GS | 522.30 | 92.86 | 0.071 | 0.656 | 0.528 | 0.838 |
| .45ACP | CB185JHP | 919.61 | 91.67 | 0.083 | 0.817 | 0.415 | 0.893 |
| .45ACP | Rem185JHP+P | 829.31 | 90.91 | 0.091 | 0.792 | 0.359 | 0.867 |
| .45ACP | Fed185JHP | 524.96 | 86.67 | 0.133 | 0.658 | 0.470 | 0.819 |
| .45ACP | Win185ST | 654.25 | 82.19 | 0.178 | 0.728 | 0.482 | 0.859 |
| .45ACP | Rem185JHP | 341.35 | 80.7 | 0.193 | 0.502 | 0.554 | 0.778 |
| .45ACP | Win230RSXT | 608.33 | 80.56 | 0.194 | 0.706 | 0.507 | 0.855 |
| .45ACP | Rem230FMJ | 251.87 | 64.75 | 0.353 | 0.390 | 0.462 | 0.672 |
| .45ACP | Fed230FMJ | 251.87 | 63.1 | 0.369 | 0.390 | 0.462 | 0.672 |
| .45ACP | Win230FMJ | 251.87 | 62.57 | 0.374 | 0.390 | 0.462 | 0.672 |
| .41Mag | Win175ST | 829.58 | 88.68 | 0.113 | 0.793 | 0.460 | 0.888 |
| .41Mag | Win210JHP | 861.24 | 82.35 | 0.177 | 0.802 | 0.592 | 0.919 |
| .41Mag | Rem210JSP | 793.24 | 80.65 | 0.194 | 0.781 | 0.544 | 0.900 |
| .41Mag | Win210SWC | 361.28 | 75.44 | 0.246 | 0.523 | 0.428 | 0.727 |
| .44Sp | Win200ST | 536.33 | 75 | 0.250 | 0.665 | 0.337 | 0.778 |
| .44Sp | Fed200LHP | 293.57 | 73.47 | 0.265 | 0.446 | 0.467 | 0.705 |
| .44Sp | Win246LRN | 331.10 | 64.79 | 0.352 | 0.491 | 0.407 | 0.698 |
| .44Sp | Rem200SWC | 422.91 | 64.71 | 0.353 | 0.582 | 0.445 | 0.768 |
| .44Mag | Win210ST | 922.22 | 90 | 0.100 | 0.817 | 0.500 | 0.909 |
| .44Mag | Fed180JHP | 1278.00 | 89.19 | 0.108 | 0.880 | 0.568 | 0.948 |
| .44Mag | Rem240SJHP | 834.88 | 88.24 | 0.118 | 0.794 | 0.566 | 0.911 |
| .45Colt | Fed225LHP | 396.74 | 78.26 | 0.217 | 0.558 | 0.595 | 0.821 |
| .45Colt | Win225ST | 598.54 | 73.58 | 0.264 | 0.701 | 0.480 | 0.844 |
| .45Colt | Win255LRN | 280.84 | 69.49 | 0.305 | 0.429 | 0.462 | 0.693 |
| .45Colt | Rem250LRN | 307.88 | 63.16 | 0.368 | 0.463 | 0.462 | 0.711 |
| .32ACP | Win71FMJ | 79.92 | 50.41 | 0.496 | 0.103 | 0.314 | 0.385 |
| .22LR | CCIsting | 218.31 | 33.92 | 0.661 | 0.340 | 0.092 | 0.401 |
| .22LR | Win37LHP | 144.66 | 28.92 | 0.711 | 0.218 | 0.098 | 0.294 |
| .22LR | Win40RNL | 120.88 | 20.97 | 0.790 | 0.175 | 0.131 | 0.283 |
| .25ACP | Win45XP | 91.36 | 25.21 | 0.748 | 0.123 | 0.203 | 0.301 |